\documentstyle[12pt,epsfig]{article}
\begin{document}

\author{L. Herrera\thanks{%
On leave from Departamento de F\'\i sica, Facultad de Ciencias, Universidad
Central de Venezuela, Caracas, Venezuela and Centro de Astrof\'\i sica
Te\'orica, M\'erida, Venezuela.} \\
%EndAName
\'Area de F\'\i sica Te\'orica\\
Facultad de Ciencias\\
Universidad de Salamanca\\
37008, Salamanca, Espa\~na.\\
and \and J. Mart\'\i nez \\
%EndAName
Grupo de F\'\i sica Estad\'\i stica\\
Departamento de F\'\i sica\\
Universidad Aut\'onoma de Barcelona\\
08193 Bellaterra, Barcelona, Espa\~na.}
\title{Dissipative fluids out of hydrostatic equilibrium}
\maketitle

\begin{abstract}
In the context of the M\"{u}ller-Israel-Stewart second order
phenomenological theory for dissipative fluids, we analyze the effects of
thermal conduction and viscosity in a relativistic fluid, just after its
departure from hydrostatic equilibrium, on a time scale of the order of
relaxation times. Stability and causality conditions are contrasted with
conditions for which the ''effective inertial mass'' vanishes.
\end{abstract}

\newpage

\section{Introduction}

The study of the evolution of self-gravitating systems (even in the
spherically symmetric case) requires the use of numerical procedures and/or
the introduction of simplifying assumptions. The former usually lead to
model dependent conclusions and the later are frequently too restrictive
and/or deprived of physical justification.

An alternative path to this question consists in perturbing the system,
compeling it to withdraw from equilibrium state. Evaluating it after its
departure from equilibrium, it is possible to study the tendency of the
evolution of the object. This is usually done following a first order
perturbative method which neglects cuadratic and higher terms in the
perturbed quantities. If the relevant processes occuring in the
self-gravitating object take place on time scales which are of the order of,
or smaller than, hydrostatic time scale, then the quasistatic approximation
fails (e.g. during the quick collapse phase preceding neutron star
formation). In this case it is necessary to evaluate the system immediately
after its departure from equilibrium, where {\it immediately} means on a
time scale of the order of relaxation times.

This approach has proven to be useful in the dissipationless case \cite
{Herrera94,Pral94,HeVa97}. Nevertheless, it has been recently found \cite
{Heal97,HeMa97} that for non-viscous dissipative case this approach cannot
always be applied. In particular, the goodness of the first order
perturbation method can be examined by means of the value of the local
parameter 
\[
\alpha =\frac{\kappa T}{\tau (\rho +p)}, 
\]
where $\kappa $ is the thermal conductivity, $T$ is the temperature, $\tau $
is the relaxation time for thermal signals and $\rho $ and $p$ are the
energy density and the radial pressure respectively. If $\alpha =1$ then it
has been shown that \cite{Heal97,HeMa97} the effective inertial mass density
of a fluid element vanishes and becomes negative if $\alpha >1$. The point
for which the system reaches condition $\alpha =1$ is called the {\it %
critical point}. This strange behaviour of matter, at, and beyond the
critical point might suggest that first order perturbation method fails
under such conditions. In some cases, systems with a value of $\alpha $
close to, or beyond, the critical point are forbidden by causality
conditions, but this is not always true. The existence of this critical
point seems to take a special relevance for the last ones. Efectively, if
causality conditions do not forbid the critical point, then there can exist
systems that cannot be studied using a first order perturbation method.
Furthermore, causality conditions \cite{HiLi83} have been found by means of
a perturbative method up to first order in the perturbed quantities. Thus,
for such systems, causality conditions must be taken with care.

The aim of this paper is to elucidate the existence of a similar critical
point in dissipative viscous systems. To do this we shall assume that,
initially, the system is either static or slowly evolving along a sequence
of states in which it is not only in hydrostatic equilibrium, but also
thermally adjusted. Then, we shall perturb the dissipative flows and radial
velocity (as seen by a Minkowskian observer). As it has been mentioned
above, the system must be evaluated on a time scale of the order of
relaxation times after the perturbation takes place. Thus, the properties of
the system are still the same, and only time derivatives of perturbed
quantities have changed appreciably, but not the quantities themselves.

In order to find the critical point we shall use the transport equations for
the dissipative flows and the radial momentum conservation equation. The
Eckart-Landau transport equations \cite{Eckart40,Landau59} imply a vanishing
relaxation time for dissipative flows. The adoption of these equations is
not advisable here for two reasons: First, they predict an infinite speed
for thermal and viscous signals propagation and unstable equilibrium states 
\cite{HiLi83}. Second, we are evaluating the system immediately after
perturbation (in the sense described above). Thus, to be consisten with this
choice we must use transport equations with non vanishing relaxation times.
In this work, transport equations are introduced using the
M\"{u}ller-Israel-Stewart second order phenomenological theory for
dissipative fluids \cite{Muller67,IsSt76}. After the condition for the
critical point is found we contrast stability and causality conditions with
this one. We shall see that, in some cases, the critical point may be
reached without violating causality conditions and, as it has been mentioned
above, causality conditions must be used with caution. Finally, we show that
neutrino trapping during gravitational collapse \cite
{Arnett77,Kazanas78,Mihalas84} can lead to values of $\alpha $ beyond the
critical point.

The paper is organized as follows. In the next section the field equations,
the conventions, and other useful formulae are introduced. In section 3 we
briefly present transport equations. In section 4 the full system of
equations is evaluated at the time when the object starts to depart from
equilibrium finding the expression for the critical point. Finally, a
discussion of the reliability of stability and causality conditions is given
in the last section.

We adopt metrics of signature -2 and geometrised units $c=G=1$ throughout
the text (except in the example preesented in last section). The quantities
subscripted with $a$ denote that they are evaluated at the surface of the
sphere.

\section{Field Equations and Conventions}

We consider spherically symmetric distributions of collapsing viscous fluid,
undergoing dissipation in the form of heat flow, and bounded by a spherical
surface $\Sigma $.

In Bondi coordinates \cite{Bondi64} the line element takes the form 
\begin{equation}
ds^2=e^{2\beta }\left[ \frac Vrdu^2+2dudr\right] -r^2\left( d\theta
^2+sin^2\theta d\varphi ^2\right) ,  \label{1}
\end{equation}
where $u=x^0$ is a timelike coordinate ($g_{uu}>0$), $r=x^1$ is a null
coordinate ($g_{rr}=0$), and $\theta =x^2$ and $\varphi =x^3$ are the usual
angle coordinates. The $u$-coordinate is the retarded time in the flat
space-time and therefore, $u-$constant surfaces, are null cones open to the
future. $V$ and $\beta $ are functions of $u$ and $r$, and the ''mass
function'', $\widetilde{m}(u,r)$, can be defined as \cite{Bondi64} 
\begin{equation}
V=e^{2\beta }(r-2\tilde{m}(u,r)).  \label{2}
\end{equation}

Bondi and Schwarzschild coordinates ($T$, $R$, $\Theta $, $\Phi $) are
related by means of the expressions 
\begin{equation}
T=u+\int_0^r\frac rVdr,  \label{3}
\end{equation}
and 
\begin{equation}
R=r,\ \ \ \Theta =\theta ,\ \ \ \Phi =\varphi .  \label{4}
\end{equation}
On the other hand, local Minkowskian coordinates ($t$, $x$, $y,$ $z$) are
related to Bondi's radiation coordinates by
\begin{equation}
dt=e^\beta \left( \sqrt{\frac Vr}du+\sqrt{\frac rV}dr\right) ,
\label{local1}
\end{equation}
\begin{equation}
dx=e^\beta \sqrt{\frac rV}dr,  \label{local2}
\end{equation}
\begin{equation}
dy=rd\theta ,  \label{local3}
\end{equation}
\begin{equation}
dz=r\sin \theta d\varphi .  \label{local4}
\end{equation}

For a local Minkowskian observer comoving with the fluid, the stress-energy
tensor splits into three terms. First, an anisotropic material part, 
\begin{equation}
\widehat{T}_{\mu \nu }^{mat}=\left( \rho ^{mat}+p_{\perp }^{mat}\right) 
\widehat{U}_\mu \widehat{U}_\nu -p_{\perp }^{mat}\eta _{\mu \nu }+\left(
p^{mat}-p_{\perp }^{mat}\right) \widehat{s}_\mu \widehat{s}_\nu  \label{4a}
\end{equation}
where , $\widehat{U}_\mu =\delta _\mu ^t$, $\widehat{s}_\mu =\delta _\mu ^x$%
, $\rho ^{mat}$ denotes the material energy density, $p^{mat}$ refers to the
material pressure for this observer, and $p_{\perp }^{mat}=p^{mat}+p_t^{mat}$
is the material part of the tangential pressure. Therefore, $p_t^{mat}$
refers to the material anisotropy. The second one is the radiation term,
which in this lagrangian frame reads \cite{Mihalas84,Lindquist66} 
\begin{equation}
\widehat{T}_{\mu \nu }^{rad}=\left( 
\begin{array}{cccc}
\rho ^{rad} & -q & 0 & 0 \\ 
-q & p^{rad} & 0 & 0 \\ 
0 & 0 & p_{\perp }^{rad} & 0 \\ 
0 & 0 & 0 & p_{\perp }^{rad}
\end{array}
\right) ,  \label{4b}
\end{equation}
where $\rho ^{rad}$ denotes the radiation energy density, $q$ the heat flow, 
$p^{rad}$ the radiation pressure, and $p_{\perp }^{rad}=\left( \rho
^{rad}-p^{rad}\right) /2$ . Finally, the viscous part can be written as 
\begin{equation}
\widehat{T}_{\mu \nu }^{vis}=\widehat{\pi }_{\mu \nu }+\Pi \widehat{h}_{\mu
\nu },  \label{4c}
\end{equation}
where $\widehat{h}_{\mu \nu }=\eta _{\mu \nu }-\widehat{U}_\mu \widehat{U}%
_\nu $ is the spatial projection tensor, and $\Pi $ is the bulk viscous
pressure. The traceless viscous pressure tensor $\widehat{\pi }_{\mu \nu }$
takes, for this observer, the form 
\begin{equation}
\widehat{\pi }_{\mu \nu }=\left( 
\begin{array}{cccc}
0 & 0 & 0 & 0 \\ 
0 & \pi & 0 & 0 \\ 
0 & 0 & -(\pi /2) & 0 \\ 
0 & 0 & 0 & -(\pi /2)
\end{array}
\right) ,  \label{4d}
\end{equation}
where $\pi $ is the shear viscous pressure.

Thus, for a local observer comoving with the fluid the stress-energy tensor
in local Minkowskian coordinates is 
\begin{equation}
\widehat{T}_{\mu \nu }=\widehat{T}_{\mu \nu }^{mat}+\widehat{T}_{\mu \nu
}^{rad}+\widehat{T}_{\mu \nu }^{vis}=(\rho +P_{\bot })\widehat{U}_\mu 
\widehat{U}_\nu -P_{\bot }\eta _{\mu \nu }+(P-P_{\bot })\widehat{s}_\mu 
\widehat{s}_\nu +2\widehat{q}_{(\mu }\widehat{U}_{\nu )},  \label{4e}
\end{equation}
where $\widehat{q}_\mu =-q\delta _\mu ^x,$ 
\begin{equation}
\rho =\rho ^{mat}+\rho ^{rad}  \label{4ee}
\end{equation}
is the total energy density, 
\begin{equation}
P=p+\pi +\Pi  \label{13}
\end{equation}
is the radial pressure, 
\begin{equation}
p=p^{mat}+p^{rad}  \label{13a}
\end{equation}
is the non-viscous radial pressure, and 
\begin{equation}
P_{\bot }=p_{\perp }^{mat}+p_{\perp }^{rad}-\frac \pi 2+\Pi =p_{\perp
}-\frac \pi 2+\Pi  \label{14}
\end{equation}
is the tangential pressure. These physical variables are obtained as
measured by this Minkowskian observer, and the effects of gravitation are
introduced by means of the local coordinate transformation ($\Lambda _\mu
^\nu $) between Minkowski coordinates and the Bondi ones (\ref{local1}-\ref
{local4}). The dynamics of the system (the radial velocity of a fluid
element as measured by a Minkowskian observer at rest in Bondi coordinates%
{\it ,} $w$) can be studied applying a Lorentz boost, $L_\mu ^\nu (-w)$, in
the radial direction to $\widehat{T}_{\mu \nu }.$ Thus, the stress-energy
tensor as measured by an observer using Bondi coordinates, with a radial
velocity, with respect to the matter configuration $-w$, is given by the
expression \cite{Bondi64} 
\begin{equation}
T_{\mu \nu }=L_\mu ^\alpha (-w)L_\nu ^\beta (-w)\Lambda _\alpha ^\gamma
\Lambda _\beta ^\delta \widehat{T}_{\gamma \delta },  \label{4f}
\end{equation}
{\it i.e.} 
\begin{equation}
T_{\mu \nu }=(\rho +P_{\bot })U_\mu U_\nu -P_{\bot }g_{\mu \nu }+(P-P_{\bot
})s_\mu s_\nu +2q_{(\mu }U_{\nu )},  \label{5}
\end{equation}
with 
\begin{equation}
s_\mu =-\frac{q_\mu }q,  \label{6}
\end{equation}
\begin{equation}
q^\mu =qe^{-\beta }\left( -\delta _u^\mu \sqrt{\frac rV}\sqrt{\frac{1-w}{1+w}%
}+\delta _r^\mu \sqrt{\frac Vr}\frac 1{\sqrt{1-w^2}}\right) ,  \label{7}
\end{equation}
and 
\begin{equation}
U^\mu =e^{-\beta }\left( \delta _u^\mu \sqrt{\frac rV}\sqrt{\frac{1-w}{1+w}}%
+\delta _r^\mu \sqrt{\frac Vr}\frac w{\sqrt{1-w^2}}\right) .  \label{8}
\end{equation}
Note that 
\begin{equation}
U^\mu q_\mu =0.  \label{nueva}
\end{equation}
The traceless viscous tensor is, for this observer 
\begin{equation}
\pi _{\mu \nu }=\left( 
\begin{array}{cccc}
e^{2\beta }\frac Vr\left( \frac{w^2}{1-w^2}\right) \pi & -e^{2\beta }\left(
\frac w{1+w}\right) \pi & 0 & 0 \\ 
-e^{2\beta }\left( \frac w{1+w}\right) \pi & e^{2\beta }\frac rV\left( \frac{%
1-w}{1+w}\right) \pi & 0 & 0 \\ 
0 & 0 & -r^2\pi /2 & 0 \\ 
0 & 0 & 0 & -r^2\sin ^2\left( \theta \right) \pi /2
\end{array}
\right) .  \label{25}
\end{equation}

Thus, Einstein equations for the line element (\ref{1}), read 
\begin{equation}
\frac{e^{-2\beta }}{8\pi }\frac rV\left[ \frac{V_{,0}-2\beta _{,0}V}{r^2}%
+\frac V{r^3}\left( e^{2\beta }-V_{,1}+2\beta _{,1}V\right) \right] =\frac{%
1\ }{1-w^2}\left( \rho +2wq+Pw^2\right) ,  \label{9}
\end{equation}
\begin{equation}
\frac{e^{-2\beta }}{8\pi }\frac 1{r^2}\left( e^{2\beta }-V_{,1}+2\beta
_{,1}V\right) =\frac 1{1+w}\left( \rho -q\left( 1-w\right) -Pw\right) ,
\label{10}
\end{equation}
\begin{equation}
\frac{e^{-2\beta }}{2\pi }\frac V{r^2}\beta _{,1}=\left( \frac{1-w}{1+w}%
\right) \left( \rho -2q+P\right) ,  \label{11}
\end{equation}
and 
\begin{equation}
e^{-2\beta }\left( 2\beta _{,01}-\frac 1{2r^2}\left[ rV_{,11}-2\beta
_{,1}V+2r\left( \beta _{,11}V+\beta _{,1}V_{,1}\right) \right] \right)
=-8\pi P_{\perp },  \label{12}
\end{equation}
where subscripts $,0$ and $,1$ denote partial derivative with respect to $u$
and $r$ coordinates respectively. From (\ref{9}) and (\ref{10}), it follows 
\begin{equation}
\frac{e^{-2\beta }}{8\pi }\frac rV\left[ \frac{V_{,0}-2\beta _{,0}V}{r^2}%
\right] =\frac{1\ }{1-w^2}\left( \rho w+q\left( 1+w^2\right) +Pw\right) .
\label{15}
\end{equation}

Next, from the conservation equation $T_{r;\mu }^\mu =0$, we obtain after
long but simple calculations 
\begin{equation}
-e^{-2\beta }\frac{\beta _{,10}}{2\pi r}+\widetilde{P}_{,1}+\frac{\left( 
\widetilde{P}+\widetilde{\rho }\right) }{1-2\widetilde{m}/r}\left[ 4\pi r%
\widetilde{P}+\frac{\widetilde{m}}{r^2}\right] -\frac 2r\left( P_{\perp
}-P\right) -\frac 2r\left( P-\widetilde{P}\right) =0,  \label{16}
\end{equation}
where 
\begin{equation}
\widetilde{P}=\frac 1{1+w}\left[ -w\rho -q\left( 1-w\right) +P\right] ,
\label{17}
\end{equation}
and 
\begin{equation}
\widetilde{\rho }=\frac 1{1+w}\left[ \rho -q\left( 1-w\right) -Pw\right] .
\label{18}
\end{equation}
Finally, taking the $u$-derivative of (\ref{11}) and using (\ref{2}), (\ref
{13}), (\ref{14}), (\ref{9}), (\ref{10}) and (\ref{15}) 
\begin{equation}
\frac{\beta _{,10}}{2\pi r}=\left( \frac{\widetilde{P}+\widetilde{\rho }}{1-2%
\widetilde{m}/r}\right) _{,0}=\frac{2r\left( 1-w\right) }{\left( 1+w\right)
(r-2\widetilde{m})}\left[ \frac{\psi _{,0}}2-\frac{2\psi w_{,0}}{1-w^2}+%
\frac{\,\psi \widetilde{m}_{,0}}{\left( r-2\,m\right) }\right] ,  \label{19}
\end{equation}
where 
\begin{equation}
\psi =p+\rho +\Pi +\pi -2q,  \label{19b}
\end{equation}
and from (\ref{9}), (\ref{2}), and (\ref{10}), we obtain 
\begin{equation}
\widetilde{m}_{,0}=-4\pi r(r-2\widetilde{m})e^{2\beta }\left[ \frac
1{1-w^2}\right] \left( \rho w+q(1+w^2)+Pw\right) .  \label{20}
\end{equation}

All these equations will be used in section 4.

Outside fluid distribution the metric is the Vaidya one \cite{Vaidya51}, a
particular case of the Bondi metric with $\beta =0$, and $V=r-2m$. The
continuity of the first and second fundamental forms across $\Sigma $ leads
to the well-known result \cite{Santos85} 
\begin{equation}
q_a=P_a,  \label{20a}
\end{equation}
or equivalently (see \cite{HeJi83} for details) 
\begin{equation}
\widetilde{P}_a=-w_a\widetilde{\rho }_a.  \label{20b}
\end{equation}

\section{Transport Equations}

As we mentioned before we shall use the M\"{u}ller-Israel-Stewart second
order phenomenological theory for dissipative fluids \cite{Muller67,IsSt76}.
Although it may be not reasonable in some situations, we shall assume here
for simplicity, that there is not viscous/heat coupling ({\it i.e.} $\alpha
_0=\alpha _1=0$ in \cite{IsSt76}). Thus, transport equations read \cite
{MaTr97} 
\begin{equation}
\tau _\kappa h_\nu ^\mu \dot{q}^\nu +q^\mu =\kappa h^{\mu \nu }\left(
T_{,\nu }-T\dot{U}_\nu \right) -\frac 12\kappa T^2\left( \frac{\tau _\kappa
U^\alpha }{\kappa T^2}\right) _{;\alpha }q^\mu +\tau _\kappa \omega ^{\mu
\nu }q_\nu ,  \label{21}
\end{equation}
\begin{equation}
\tau _\zeta \dot{\Pi }+\Pi =-\zeta \Theta -\frac 12\zeta T\left( \frac{\tau
_\zeta U^\alpha }{\zeta T}\right) _{;\alpha }\Pi ,  \label{22}
\end{equation}
and 
\begin{equation}
\tau _\eta h_\mu ^\alpha h_\nu ^\beta \dot{\pi }_{\alpha \beta }+\pi _{\mu
\nu }=2\eta \sigma _{\mu \nu }-\frac 12\eta T\left( \frac{\tau _\eta
U^\alpha }{\eta T}\right) _{;\alpha }\pi _{\mu \nu }+2\tau _\eta \pi _{(\mu
}^\alpha \omega _{\mu )\alpha },  \label{23}
\end{equation}
where $h^{\mu \nu }$ is the projector onto the three space orthogonal to $%
U^\mu $, $\omega _{\mu \nu }=h_\mu ^\alpha h_\nu ^\beta U_{[\alpha ;\beta ]}$
is the vorticity, $\Theta =U_{;\mu }^\mu $ is the expansion scalar, and $%
\kappa $, $\zeta $, and $\eta $ denote the thermal conductivity, and the
bulk and shear viscous coefficients respectively. Also, $T$, $\tau _\kappa $%
, $\tau _\zeta $, and $\tau _\eta $ denote temperature and relaxation times
respectively. Overdot denotes $\dot{A}_{\alpha \beta \ldots }=U^\lambda
A_{\alpha \beta \ldots ;\lambda }$, and the shear tensor is given by 
\begin{equation}
\sigma _{\mu \nu }=h_\mu ^\alpha h_\nu ^\beta U_{(\alpha ;\beta )}-\frac
13h_{\mu \nu }h^{\alpha \beta }U_{\alpha ;\beta }.  \label{24}
\end{equation}
The traceless viscous pressure tensor $\pi _{\mu \nu }$ is given by the
expression (\ref{25}).

Observe that, due to the symmetry of the problem, equations (\ref{21}), and (%
\ref{23}) only have one independent component \cite{StEl68}.

Let us now write the expressions for different terms in (\ref{21}), they are 
\begin{eqnarray}
\tau _\kappa h_\nu ^r\dot{q}^\nu  &=&\tau _\kappa \frac{e^{-2\,\beta }}{%
\left( 1+w\right) }\left( q_{,0}+\frac Vr\frac w{1-w}q_{,1}\right) +\frac{%
\tau _\kappa qwe^{-2\beta }}{1-w}\left( \frac Vr\right) \stackrel{.}{U}_r 
\nonumber \\
&&+\tau _\kappa qw\frac{e^{-2\beta }}{(1-w^2)(1+w)}\left( w_{,0}+\frac
Vr\frac w{1-w}w_{,1}\right)   \nonumber \\
&&-\frac{\tau _\kappa qw}{1-w^2}\left( -2\beta _{,1}\left[ 1-\frac{2%
\widetilde{m}}r\right] +\frac{\widetilde{m}_{,1}}r-\frac{\widetilde{m}}{r^2}+%
\frac{\widetilde{m}_{,0}}V(1-w)\right)   \label{26}
\end{eqnarray}
\begin{equation}
q^r=qe^{-\beta }\sqrt{\frac Vr}\frac 1{\sqrt{1-w^2}},  \label{27}
\end{equation}
\begin{equation}
\kappa h^{r\nu }T_{,\nu }=\frac{\kappa e^{-2\,\beta }}{1+w}\left(
T_{,0}-\frac Vr\frac{T_{,1}}{1-w}\right) ,  \label{28}
\end{equation}
\begin{equation}
-\kappa Th^{r\nu }\dot{U}_\nu =\kappa Te^{-2\beta }\frac Vr\left[ \frac
1{1-w}\right] \dot{U}_r,  \label{29}
\end{equation}
\[
-\frac 12\kappa T^2\left( \frac{\tau _\kappa U^\alpha }{\kappa T^2}\right)
_{;\alpha }q^r=-\frac{\tau _\kappa }2\Theta q^r
\]
\begin{equation}
-\frac 12\kappa T^2q\frac{e^{-2\beta }}{1+w}\left[ \left( \frac{\tau _\kappa 
}{\kappa T^2}\right) _{,0}+\frac Vr\left( \frac w{1-w}\right) \left( \frac{%
\tau _\kappa }{\kappa T^2}\right) _{,1}\right] ,  \label{30}
\end{equation}
and 
\begin{equation}
\tau _\kappa \omega ^{r\nu }q_\nu =\tau _\kappa h^{r\alpha }h^{\nu \beta
}U_{[\alpha ;\beta ]}q_\nu =\tau _\kappa h^{r\alpha }U_{[\alpha ;\beta
]}q^\beta =0,  \label{nueva2}
\end{equation}
where 
\[
\dot{U}_r=\frac 1{1+w}\left( \frac 1{2r}-\beta _{,1}-\frac{V_{,1}}{2V}%
\right) +\frac rV\left( \frac{1-w}{1+w}\right) \left( \beta _{,0}-\frac{%
V_{,0}}{2V}\right) 
\]
\begin{equation}
-\frac 1{\left( 1+w\right) ^2\left( 1-w\right) }\left( ww_{,1}+\frac
rV(1-w)w_{,0}\right) ,  \label{31}
\end{equation}
and the expansion scalar is given by 
\[
\Theta =e^{-\beta }\sqrt{\frac Vr}\frac{2w}{\sqrt{1-w^2}}\left( \frac
1r+\beta _{,1}\right) +e^\beta \sqrt{\frac rV}\frac w{\sqrt{1-w^2}}\left( 
\frac{\widetilde{m}}{r^2}-\frac{\widetilde{m}_{,1}}r\right) 
\]
\begin{equation}
+\frac{e^\beta }r\left( \frac rV\right) ^{3/2}\widetilde{m}_{,0}\sqrt{\frac{%
1-w}{1+w}}+e^{-\beta }\sqrt{\frac Vr}\frac 1{\sqrt{1-w^2}\left( 1+w\right)
}\left[ \frac{w_{,1}}{1-w}-\frac rVw_{,0}\right] .  \label{36}
\end{equation}

For (\ref{22}), we have 
\begin{equation}
\tau _\zeta \dot{\Pi}=\tau _\zeta U^\alpha \Pi _{,\alpha }=\tau _\zeta
e^{-\beta }\sqrt{\frac rV}\sqrt{\frac{1-w}{1+w}}\left[ \Pi _{,0}+\frac
Vr\frac w{1-w}\Pi _{,1}\right] ,  \label{34}
\end{equation}
and 
\[
-\zeta \Theta -\frac 12\zeta T\left( \frac{\tau _\zeta U^\alpha }{\zeta T}%
\right) _{;\alpha }\Pi =-\zeta \Theta \left( 1+\tau _\zeta \frac \Pi
2\right) 
\]
\begin{equation}
-\frac 12\Pi \zeta Te^{-\beta }\sqrt{\frac rV}\sqrt{\frac{1-w}{1+w}}\left[
\left( \frac{\tau _\zeta }{\zeta T}\right) _{,0}+\frac Vr\frac w{1-w}\left( 
\frac{\tau _\zeta }{\zeta T}\right) _{,1}\right] .  \label{35}
\end{equation}

Finally, for the different terms in (\ref{23}) we get 
\begin{equation}
\pi _{\theta \theta }=-\frac{r^2}2\pi ,  \label{37}
\end{equation}
\begin{equation}
\tau _\eta h_\theta ^\alpha h_\theta ^\beta \dot{\pi}_{\alpha \beta }=\tau
_\eta U^\alpha \dot{\pi}_{\theta \theta ;\alpha }=-\frac{r^2}2e^{-\beta
}\tau _\eta \sqrt{\frac rV}\sqrt{\frac{1-w}{1+w}}\left[ \pi _{,0}+\frac
Vr\frac w{1-w}\pi _{,1}\right] ,  \label{38}
\end{equation}
\begin{equation}
2\eta \sigma _{\theta \theta }=2\eta \left[ -e^{-\beta }\sqrt{\frac Vr}\frac
w{\sqrt{1-w^2}}r+\frac \Theta 3r^2\right] ,  \label{39}
\end{equation}
\[
-\frac 12\eta T\left( \frac{\tau _\eta U^\alpha }{\eta T}\right) _{;\alpha
}\pi _{\theta \theta }=\frac{r^2}4\pi \tau _\eta \Theta 
\]
\begin{equation}
+\frac{r^2}4\pi \eta Te^{-\beta }\sqrt{\frac rV}\sqrt{\frac{1-w}{1+w}}\left[
\left( \frac{\tau _\eta }{\eta T}\right) _{,0}+\frac Vr\frac w{1-w}\left( 
\frac{\tau _\eta }{\eta T}\right) _{,1}\right] ,  \label{40}
\end{equation}
and 
\begin{equation}
2\tau _\eta \pi _{(\theta }^\alpha \omega _{\theta )\alpha }=0.
\end{equation}

Note that for this observer the shear scalar and vorticity scalar are given
by 
\begin{equation}
\sigma =\sqrt{\frac 12\sigma _{\mu \nu }\sigma ^{\mu \nu }}=\frac{\sqrt{3}}{%
r^2}\left( \frac \Theta 3r^2-e^{-\beta }\sqrt{\frac Vr}\frac w{\sqrt{1-w^2}%
}\right) ,
\end{equation}
and 
\begin{equation}
\frac 12\omega _{\mu \nu }\omega ^{\mu \nu }=h^{\alpha \gamma }h^{\beta
\delta }U_{[\alpha ;\beta ]}U_{[\gamma ;\delta ]}=0,
\end{equation}
respectively.

Transport equations (\ref{21}-\ref{23}), together with (\ref{16}) will be
evaluated after the system departs from equilibrium, neglecting terms of
order ${\cal O}(w ^2)$ and higher.

\section{Departure from Hydrostatic Equilibrium}

We assume that, before perturbation, the system is slowly evolving along a
sequence of states in which it is close to hydrostatic equilibrium and
thermally adjusted - the so called {\it complete equilibrium }\cite[p.66]
{KiWe94}. Thus, the radial velocity, as seen by a Minkowskian observer, is
small. This means that cuadratic and higher terms in $w$ may be neglected in
a first order perturbation theory. A system is thermally adjusted if it
changes its properties considerabily only within a time scale $\tau _{cha}$
that is large as compared with the Kelvin-Helmholtz time scale $\tau _{KH}$.
Thus, before perturbation we can assume that the $u$-derivatives of the
perturbed quantities can be neglected up to first order, and consequently 
\begin{equation}
q_{,0}\sim \Pi _{,0}\sim \pi _{,0}\sim w_{,0}\sim {\cal O}(w^2).
\label{aprox}
\end{equation}
On the other hand, the hydrostatic equilibrium can be justified in terms of
the characteristic times: If the hydrostatic time scale $\tau _{hyd}\sim 
\sqrt{r^3/m}$ is much shorter than the Kelvin-Helmholtz time scale $\tau
_{KH}\sim m^2/2rl,$ then inertial terms in the equation of motion $T_{r;\mu
}^\mu =0$ can be ignored. This condition will be accomplished for small
values of luminosity $l$, and consequently for small values of $q.$ Thus,
before perturbation, we can assume $q\sim {\cal O}(w)$ in the whole system.
It seems also reasonable to assume that in such system, bulk viscous
pressure and shear viscous pressure must be also small ({\it i.e. }$\Pi \sim
\pi \sim {\cal O}(w)$).

Note that from (\ref{15}) and (\ref{20}) $\beta _{,0}$, $V_{,0}$ and $m_{,0}$
are of order $w.$ Therefore, their products and second time derivatives may
be neglected (${\cal O}(w^n)$; $n\geq 2$) (an invariant characterization of
slow evolution may be found in \cite{HeSa95}).

We shall evaluate the system immediately after perturbation (in the sense
described in the introduction). Physically, this implies that the perturbed
quantities ($\omega ,$ $q,$ $\pi $ and $\Pi $) are still much less than
unity. Nevertheless, the system is departing form hydrostatic equilibrium
and thermal adjustment. Thus, the $u$-derivatives of the perturbed
quantities are small but different from zero ({\it i.e.} $q_{,0}\sim
w_{,0}\sim \pi _{,0}\sim \Pi _{,0}\sim {\cal O}(w)$).

Thus, our initially slowly evolving system is characterized by:

\begin{enumerate}
\item  Before perturbation 
\begin{equation}
\begin{array}{c}
\rho _{,0}\approx p_{,0}\approx p_{\perp ,0}\approx w\approx q\approx \pi
\approx \Pi \approx \widetilde{m}_{,0}\approx {\cal O}(w) \\ 
w_{,0}\approx q_{,0}\approx \pi _{,0}\approx \Pi _{,0}\approx {\cal O}(w^2)
\end{array}
\label{41}
\end{equation}

\item  After perturbation 
\begin{equation}
\begin{array}{c}
\rho _{,0}\approx p_{,0}\approx p_{\perp ,0}\approx w\approx q\approx \pi
\approx \Pi \approx \widetilde{m}_{,0}\approx {\cal O}(w) \\ 
w_{,0}\approx q_{,0}\approx \pi _{,0}\approx \Pi _{,0}\approx {\cal O}(w)
\end{array}
\label{42}
\end{equation}
\end{enumerate}

In both cases we have 
\begin{equation}
\widetilde{P}=P-(wp+w\rho +q)+{\cal O}(w^2)  \label{43}
\end{equation}
\begin{equation}
\widetilde{\rho }=\rho -(wp+w\rho +q)+{\cal O}(w^2)  \label{44}
\end{equation}
The initially static case can also be considered. This system is
characterized by:

\begin{enumerate}
\item  Before perturbation 
\begin{equation}
\begin{array}{c}
w=q=\pi =\Pi =\widetilde{m}_{,0}=0 \\ 
\rho _{,0}=p_{,0}=p_{\perp ,0}=w_{,0}=q_{,0}=\pi _{,0}=\Pi _{,0}=0.
\end{array}
\label{44bis}
\end{equation}

\item  After perturbation 
\begin{equation}
\begin{array}{c}
w=q=\pi =\Pi =\widetilde{m}_{,0}=\rho _{,0}=p_{,0}=p_{\perp ,0}=0 \\ 
w_{,0}\approx q_{,0}\approx \pi _{,0}\approx \Pi _{,0}\neq 0%
\mbox{\qquad
(small).}
\end{array}
\label{44tris}
\end{equation}
\end{enumerate}

And 
\begin{equation}
\widetilde{P}=P,\qquad \widetilde{\rho }=\rho .
\end{equation}
Let us now start by evaluating (\ref{21}). In the static case we obtain
before perturbation 
\begin{equation}
\frac{T_{,1}}T=\frac 1{2r}-\frac{V_{,1}}{2V}-\beta _{,1},  \label{45}
\end{equation}
and immediately after perturbation, neglecting terms of order ${\cal O}(w^2)$
and higher, 
\begin{equation}
\tau _\kappa q_{,0}=-\kappa Tw_{,0}+{\cal O}(w^2).  \label{47}
\end{equation}
For bulk viscous pressure equation, we obtain in the initially static case
after perturbation 
\begin{equation}
\tau _\zeta \Pi _{,0}=\zeta w_{,0}+{\cal O}(w^2).  \label{48}
\end{equation}
Expressions (\ref{47}) and (\ref{48}) are also obtained for the initially
slowly evolving case applying (\ref{41}) and (\ref{42}).

The evaluation of the equation for the shear viscous pressure (\ref{23})
yields, after perturbation, for both possible initial configurations 
\begin{equation}
\tau _\eta \pi _{,0}=\frac 43\eta w _{,0}+{\cal O}(w ^2).  \label{49}
\end{equation}
Thus, the three transport equations (\ref{21}-\ref{23}), evaluated after
perturbation, lead to expressions which are the same for the two initial
configurations.

Finally, let us evaluate conservation equation $T_{r;\mu }^\mu =0$ (\ref{16}%
) after perturbation. In the initially static case, we have before
perturbation condition (\ref{44bis}). Therefore, equation (\ref{16}) becomes 
\begin{equation}
R\equiv P_{,1}+\frac{\left( P+\rho \right) }{1-2m/r}\left[ 4\pi rP+\frac
m{r^2}\right] -\frac 2r\left( p_{\perp }-p\right) =0,  \label{54}
\end{equation}
which is the equation of hydrostatic equilibrium for anisotropic fluids, and 
$-R$ denotes the total outward force acting on a given fluid element. After
perturbation we obtain, using (\ref{19}), (\ref{20}), and (\ref{47}-\ref{49}%
) 
\begin{equation}
-R=\frac{2e^{-2\beta }(\rho +p)}{(1-2m/r)}\left[ 1-\alpha \right] \times
w_{,0},  \label{54b}
\end{equation}
where 
\begin{equation}
\alpha =\frac 1{(\rho +p)}\left( \frac \zeta {2\tau _\zeta }+\frac{2\eta }{%
3\tau _\eta }+\frac{\kappa T}{\tau _\kappa }\right) ,  \label{61}
\end{equation}
or equivalently 
\begin{equation}
w_{,0}=-\left[ \frac{e^{2\beta }R}2\right] \frac{(1-2m/r)}{(\rho +p)}\times
\left[ 1-\alpha \right] ^{-1}.  \label{55}
\end{equation}

Assuming the second initial case (slowly evolving), conservation equation $%
T_{r;\mu }^\mu =0$ with (\ref{19}), (\ref{20}), and conditions (\ref{47}-\ref
{49}) lead, before perturbation, to expression 
\begin{equation}
F=-\frac{e^{-2\beta }}{(r-2\widetilde{m})}\left[ r\left( p+\rho \right)
_{,0}-8\pi r^2\left( \rho +\,p\right) e^{2\beta }(\rho w+q+pw)\right] +%
\widetilde{R}-\frac 2r\left( P-\widetilde{P}\right) ,  \label{56}
\end{equation}
where 
\begin{equation}
\widetilde{R}=\widetilde{P}_{,1}+\frac{\left( \widetilde{P}+\widetilde{\rho }%
\right) }{1-2\widetilde{m}/r}\left[ 4\pi r\widetilde{P}+\frac{\widetilde{m}}{%
r^2}\right] -\frac 2r\left( P_{\perp }-P\right) ,  \label{57}
\end{equation}
and $-F$ (as well as $-R$ in the previous case) may be easily interpreted as
the total outward force acting on a given fluid element. After perturbation
we obtain from (\ref{19}) and (\ref{20}) 
\[
\frac{\beta _{,10}}{2\pi r}=\frac r{(r-2\widetilde{m})}\times 
\]
\begin{equation}
\left[ \left( p+\rho +\Pi +\pi -2q\right) _{,0}-2\left( \rho +p\right)
\left( w_{,0}+4\pi re^{2\beta }(\rho w+q+pw)\right) \right] .  \label{58}
\end{equation}
Using this last expression together with (\ref{47}-\ref{49}), we obtain for
equation (\ref{16}) after perturbation 
\begin{equation}
-F=\frac{2e^{-2\beta }(\rho +p)}{(1-2\widetilde{m}/r)}\left[ 1-\alpha
\right] \times w_{,0},  \label{59}
\end{equation}
or 
\begin{equation}
w_{,0}=-\left[ \frac{e^{2\beta }F}2\right] \frac{(1-2\widetilde{m}/r)}{(\rho
+p)}\times \left[ 1-\alpha \right] ^{-1}.  \label{60}
\end{equation}

Equations (\ref{55}) and (\ref{60}) may be compared with the Newtonian form %
\centerline{Force $=$ mass $\times$ acceleration,} where here the term

\begin{equation}
\frac{2e^{-2\beta }(\rho +p)}{(1-2\widetilde{m}/r)}\left[ 1-\alpha \right] ,
\end{equation}
stands for the {\it effective inertial mass }. This one vanishes for $\alpha
=1$, implying the vanishing of $-F$, even though the time derivative of the
radial velocity is different from zero. As it has been mentioned in the
introduction, $\alpha =1$ corresponds to the {\it critical point. }This one
coincides with the given in \cite{Heal97} if $\zeta $ and $\eta $ are zero.

Note that the effective inertial mass decreases as $\alpha $ grows. As $%
\alpha $ approaches to unity the system seems to be more unstable, and for $%
\alpha \sim 1$ a vanishingly small radial force leads to non zero values of $%
w_{,0}.$ This fact contradicts the assumption that the hydrostatic
equilibrium corresponds to a vanishing total radial force , and consequently
the reliability of a perturbative approach is in question under such
condition. This approach also predicts an anomalous behaviour beyond the
critical point. If $\alpha >1,$ then an outward force ($-F>0$) implies an
inward acceleration ($w_{,0}<0$). Thus, we may conclude that we can neglect
cuadratic and higher terms in the perturbed variables only if $\alpha $ is
not close to, or beyond, the critical point.

Causality and stability conditions \cite{HiLi83} have been found using a
perturbative method up to first order. Therefore, it seems interesting to
answer to the following question. Are systems with $\alpha \sim 1$, or $%
\alpha >1,$ always forbidden by causality conditions? If the answer is no,
then the reliability of causality conditions is uncertain for such systems.
In the next section we shall try to answer this question.

\section{Discussion}

In order to contrast the $\alpha \sim 1,$ and $\alpha >1$ conditions with
stability and causality conditions, it is convenient to write (\ref{61}),
using the notation adopted in \cite{HiLi83}. Thus, 
\begin{equation}
\alpha =\frac 1{(\rho +p)}\left( \frac 1{2\beta _0}+\frac 1{3\beta _2}+\frac
1{\beta _1}\right)  \label{62}
\end{equation}

According to linear perturbation theory \cite{HiLi83}, causality and
stability requires 
\begin{equation}
(\rho +p)(1-c_s^2)>\frac 1{\beta _0}+\frac 2{3\beta _2}+\frac{nTc_vK^2}{%
\beta _1nTc_v-1},  \label{63}
\end{equation}
\begin{equation}
(\rho +p)>\frac{2\beta _2+\beta _1}{2\beta _2\beta _1},  \label{64}
\end{equation}
and 
\begin{equation}
\beta _1>\frac 1{nTc_v},  \label{65}
\end{equation}
where 
\begin{equation}
K=1-\frac{\alpha _p}{nc_v\kappa _T}.  \label{66}
\end{equation}
The adiabatic contribution to the speed of sound is denoted by $c_s$, $n$ is
the particle number density, and $c_v$, $\kappa _T$, and $\alpha _p$ denote
specific heat at constant volume, isothermal compressibility, and thermal
expansion coefficient respectively. As usually, they are defined by 
\begin{equation}
c_v=T\left( \frac{\partial s}{\partial T}\right) _n,  \label{67}
\end{equation}
\begin{equation}
\kappa _T=\frac 1n\left( \frac{\partial n}{\partial p}\right) _T,  \label{68}
\end{equation}
and 
\begin{equation}
\alpha _p=-\frac 1n\left( \frac{\partial n}{\partial T}\right) _p.
\label{69}
\end{equation}

As we already mentioned, if the two viscosity coefficients vanishes, we
recover the result found in \cite{Heal97}. In this case it can be shown that
the critical point is very close to the point where (\ref{63}-\ref{65})
break down \cite{HeMa97} for small values of $c_s^2$.

Let us now consider the case where there is only bulk viscosity ($\kappa
=\eta =0$). In this case $\beta _1,\beta _2\rightarrow \infty $ \cite
{Maartensastr}, and the critical point is overtaken if 
\begin{equation}
\beta _0(\rho +p)<\frac 12,  \label{70}
\end{equation}
is satisfied, whereas causality requires 
\begin{equation}
\beta _0(\rho +p)>\frac 1{1-c_s^2}.  \label{71}
\end{equation}
Therefore, it appears that the critical point is forbidden by causality and
stability requirements.

In the pure shear viscosity case ($\kappa =\zeta =0$), the critical point is
overtaken if 
\begin{equation}
\beta _2(\rho +p)<\frac 13,  \label{72}
\end{equation}
whereas the most restrictive causality condition, is given (in this case) by
(\ref{63}) 
\begin{equation}
\beta _2(\rho +p)>\frac 2{3(1-c_s^2)}.  \label{73}
\end{equation}
Again, the critical point is beyond the point where causality is violated.

Let us now consider the general case ($\kappa ,\zeta ,\eta >0$). We may
write (\ref{63}) as 
\begin{equation}
\frac 1{2\beta _0}+\frac 1{3\beta _2}+\frac 1{\beta _1}<(\rho +p)\left( 
\frac{1-c_s^2}2\right) +\frac 1{\beta _1}-\frac{nTc_vK^2}{2\beta _1nTc_v-2}.
\label{74}
\end{equation}
We are going to find conditions for which the point where causality is
violated is beyond the critical point. Assuming that our system is at, or
beyond, the critical point ($\alpha \geq 1$), then we should demand by
virtue of (\ref{62}) and (\ref{74}) 
\begin{equation}
(\rho +p)\leq \frac 1{2\beta _0}+\frac 1{3\beta _2}+\frac 1{\beta _1}<(\rho
+p)\left( \frac{1-c_s^2}2\right) +\frac 1{\beta _1}-\frac{nTc_vK^2}{2\beta
_1nTc_v-2},  \label{75}
\end{equation}
or, after some elementary algebra 
\begin{equation}
\beta _1(\rho +p)<\frac{\rho +p}{nTc_v}+\left[ 1-\frac 1{\beta _1nTc_v}-%
\frac{K^2}2\right] \left( \frac 2{1+c_s^2}\right) ,  \label{76}
\end{equation}
and combining it with (\ref{65}) 
\begin{equation}
\frac{\rho +p}{nTc_v}<\beta _1(\rho +p)<\frac{\rho +p}{nTc_v}+\left[ 1-\frac
1{\beta _1nTc_v}-\frac{K^2}2\right] \left( \frac 2{1+c_s^2}\right) ,
\label{77}
\end{equation}
implying 
\begin{equation}
\left[ 1-\frac 1{\beta _1nTc_v}-\frac{K^2}2\right] \left( \frac
2{1+c_s^2}\right) >0,  \label{78}
\end{equation}
which is equivalent to 
\begin{equation}
\beta _1>\frac 1{nTc_v}\left[ \frac 2{2-K^2}\right] .  \label{79}
\end{equation}
Thus, if $K^2<2$, it is, in principle, possible to attain the critical point
without violating causality conditions (\ref{63}) and (\ref{65}). Note that
condition (\ref{64}) has not been used in this calculus, so it should be
demanded in addition to condition $K^2<2$. Therefore, from (\ref{64}), (\ref
{66}), and (\ref{79}), the critical point can be overtaken without violating
causality conditions if the following conditions are satisfied 
\begin{eqnarray}
\left( 1-\frac{\alpha _p}{nc_v\kappa _T}\right) ^2 &<&2  \label{79b} \\
\frac{\kappa T}{\tau _\kappa (\rho +p)}+\frac \eta {\tau _\eta (\rho +p)}
&<&1  \label{79c}
\end{eqnarray}
Note that conditions (\ref{79b}) and (\ref{79c}) do not imply necessarily $%
\alpha >1$, but if causality conditions and $\alpha >1$ are accomplished,
then conditions (\ref{79b}) and (\ref{79c}) must be fulfilled. An example of
this situation is an ultrarelativistic monoatomic ideal gas ($\rho \approx
3nkT/2$, $p\approx nkT$, $\gamma =mc^2/k_BT\ll 1$). For this fluid $\alpha
_p=T^{-1}$, $c_v=3k/2$, $\kappa _T=p^{-1}$, $\kappa =4T^{-1}p\tau _\kappa /5$%
, $\zeta =$ $\gamma ^4p\tau _\zeta /216$ and $\eta =2p\tau _\eta /3.$ Then (%
\ref{79b}) and (\ref{79c}) are accomplished, but $\alpha \sim 0.5$.

On the other hand if $K^2>2$, then the critical point is less restrictive
than (\ref{65}), and causality and stability conditions can be used freely.
Thus, we face the following alternatives:

\begin{enumerate}
\item  In the case of pure bulk or shear viscosity (without heat
conduction), the critical point is well beyond the point where causality
breaks down. Therefore, the system should not reach the critical point in
those cases, and linear approximation can be applied to find stability and
causality conditions.

\item  In the non-viscous case, the critical point is very close to the
point where causality is violated for small values of the sound speed. Since
the linear approximation is not reliable close to the critical point, then
it might be possible for a given system to attain the critical point.

\item  In the general case it may happen than causality breaks down beyond
the critical point. Thus, it appear that there exist situations where a
given physical system may attain the critical point and even go beyond it.
\end{enumerate}

\begin{figure}[t]
\epsfig{file=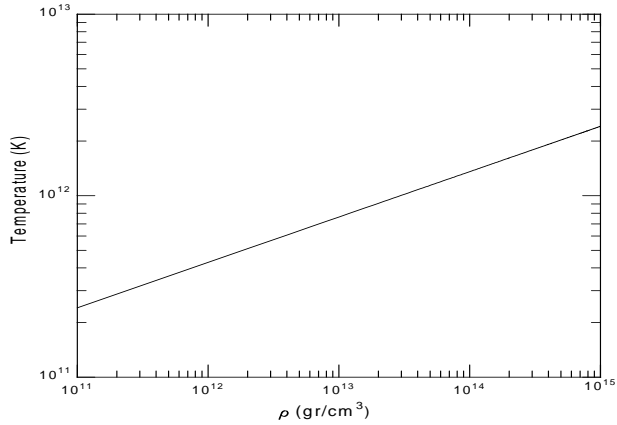, height=9.0cm, width=9.0cm}
\caption{Temperature for which $\alpha =1$ as a function of energy %
density. Systems with $\alpha >1$, are above the line.}
\end{figure}

It is worth noticing that condition $\alpha >1$ can be accomplished in non
very exotic systems. One of them is an interacting mixture of matter and
neutrinos, which is a well-known scenario during the formation of a neutron
star in a supernova explosion. In this case the heat conductivity
coefficient is given by \cite{Weinberg71,Shapiro89} 
\begin{equation}
\kappa =\frac 43bT^3\tau ,  \label{79d}
\end{equation}
where $\tau $ is the mean collision time, $b=7N_\nu a/8$, $N_\nu $ is the
number in neutrino flavors and $a$ is the radiation constant. Assuming that
two viscosity coefficients vanishes, and $p\ll\rho $ then 
\begin{equation}
\alpha =\frac{\kappa T}{\tau _\kappa \left( \rho +p\right) }\simeq \frac{%
\kappa T}{\tau \rho }.  \label{79e}
\end{equation}
Using usual units, the critical point is overtaken if 
\begin{equation}
T>\sqrt[4]{\frac{6\rho c^3}{7N_\nu a}}\sim 4.29\times 10^8\rho ^{1/4},
\label{79f}
\end{equation}
where we have adopted $\tau \sim \tau _\kappa $, $N_\nu =3$, $T$ is in
Kelvin and $\rho $ is given in g cm$^{-3}.$ The values of temperature, for
which $\alpha =1$, are presented in figure 1 as a function of energy
density. These ones are similar to the expected temperature that can be
reached during hot collapse in a supernova explosion \cite[$\S$ 18.6]{ShTe83}.

We would like to conclude with the following comment: for degenerate
matter,when thermal conductivity is dominated by electrons, thermal
relaxation time may be of the order of milliseconds (or even larger), due to
larger mean free path of electrons \cite{HeFa95} , but this is of the same
order of magnitude as the time scale of the quick phase preceding neutron
star formation. Therefore for this last scenario (at least) , the basic
assumption of our approach is justified.

\section*{Acknowledgments}

This work has been partially supported by the Spanish Ministry of Education
under Grant No. PB94-0718

\end{document}